\documentclass[floatfix,twocolumn,showpacs,preprintnumbers,amsmath,amssymb,prb]{revtex4}

\usepackage{graphicx}
\usepackage{dcolumn}
\usepackage{bm}

\begin{document}

\preprint{}

\title{Dual Geometric Worm Algorithm for Two-Dimensional Discrete Classical Lattice Models}

\author{Peter Hitchcock}
\email{hitchpa@muss.cis.mcmaster.ca}
\affiliation{Department of Physics and Astronomy, McMaster University, Hamilton,
ON, L8S 4M1 Canada
}
\author{Erik~S.~S\o rensen}%
\email{sorensen@mcmaster.ca}
\homepage{http://ingwin.physics.mcmaster.ca/~sorensen}
\affiliation{Department of Physics and Astronomy, McMaster University, Hamilton,
ON, L8S 4M1 Canada }
\author{Fabien Alet$^{a,b}$}
\email{alet@phys.ethz.ch}
\affiliation{$^a$Computational Laboratory, ETH Z\"urich, CH-8092 Z\"urich, Switzerland}
\affiliation{$^b$Theoretische Physik, ETH Z\"urich, CH-8093 Z\"urich, Switzerland}


\date{\today}

\begin{abstract}
We present a dual geometrical worm algorithm for two-dimensional Ising models.
The existence of such dual algorithms was first pointed out by Prokof'ev and
Svistunov~\cite{ProkofevClassical}.  The algorithm is defined on the dual
lattice and is formulated in terms of bond-variables and can therefore be
generalized to other two-dimensional models that can be formulated in terms of
bond-variables. We also discuss two related algorithms formulated on the direct
lattice, applicable in any dimension. These latter algorithms turn out to be
less efficient but of considerable intrinsic interest.  We show how such
algorithms quite generally can be ``directed" by minimizing the probability for
the worms to erase themselves.  Explicit proofs of detailed balance are given
for all the algorithms.  In terms of computational efficiency the dual
geometrical worm algorithm is comparable to well known cluster algorithms such
as the Swendsen-Wang and Wolff algorithms, however, it is quite different in
structure and allows for a very simple and efficient implementation.  The dual
algorithm also allows for a very elegant way of calculating the domain wall
free energy.

\end{abstract}

\pacs{05.10.Ln, 05.20.-y, O2.70.Tt}
\maketitle

\section{Introduction}
Over the recent decades many powerful Monte Carlo (MC) algorithms have been
developed, greatly enhancing the scope and applicability of Monte Carlo techniques. In
fact, it is quite likely that these algorithmic advances have, and will
continue to have, a far greater impact on the predictive power of Monte Carlo
simulations than advances in raw computational capacity, a point that is often
overlooked. The continued development of such advanced algorithms is therefore very important.
Here we shall mainly be concerned with MC algorithms suitable for the study of lattice models
described by classical statistical mechanics. Some of the most notable developments in this field
have been the development of cluster algorithms by Swendsen and Wang~\cite{SwendsenWang} and
by Wolff~\cite{Wolff}. More recent developments include invaded cluster algorithms~\cite{machta}
that self-adjust to the critical temperature, flat histogram methods~\cite{flathistogram}, focusing
on the density of states and techniques performing Monte Carlo sampling of the high temperature
series expansion of the partition function~\cite{ProkofevClassical} using worm algorithms~\cite{prokofev}.
Two of us recently proposed a very efficient geometrical
worm algorithm~\cite{bosons,directed} for the bosonic Hubbard model. In this algorithm variables are
not updated at random but instead a ``worm" is propagated through the lattice, at each step choosing
a new site to visit with a probability proportional to the {\it relative} (among the different sites
considered) probability of changing
the corresponding variable. The high efficiency of the algorithm stems from the fact that the worm is {\it always} moved
and it is only through the actual movement that
the local environment - the local ``geometry" - comes into play. The ideas underlying this algorithm
are quite generally applicable and the present paper is concerned
with their generalization to classical statistical mechanics models.

As the canonical testing ground for new algorithms we consider the standard ferromagnetic Ising model
in two dimensions defined by:
\begin{equation}
H=-J\sum_{<i,j>}s_is_j,\ \ s_i=\pm 1.
\label{eq:Ising}
\end{equation}
Here $<i,j>$ denote the summation over nearest neighbor spins.
It is well known that the critical temperature for this model in two dimensions is
$k_B T_c=2J/\log(1+\sqrt{2})=J2.26918\ldots$. When investigating magnetic materials modeled by
classical statistical mechanics such as Eq.~(\ref{eq:Ising}) using Monte Carlo methods one has
to take into account the effects of the non-zero auto-correlation time $\tau$ (defined below) that is always
present in Monte Carlo simulations. The auto-correlation time describes the correlation between observations
of an observable ${\cal O}(t_0)$ and ${\cal O}(t_0+t)$, $t$ Monte Carlo Sweeps (MCS) apart. The auto-correlation 
time, $\tau$, depends on the simulation temperature and the system size and grows dramatically close to the
critical temperature, $T_c$, a phenomenon referred to as critical slowing down. At $T_c$ the auto-correlation 
time displays a power-law dependence of the system size, $\tau\sim L^{z_{MC}}$, defining a Monte Carlo
dynamical exponent $z_{MC}$. For the well known Metropolis algorithm one estimates~\cite{Newman,Landau}
$z_{MC}\sim 2.1-2.2$. If efficient algorithms, with a very small $z_{MC}$, cannot be found, this scaling
renders the Monte Carlo method essentially useless for large lattice sizes since all data will be correlated.
It is therefore crucial to develop algorithms with a very small or zero $z_{MC}$. In order to test 
the proposed algorithms we shall therefore only consider simulations at the critical point since this is
where the critical slowing down is the most pronounced and where $z_{MC}$ is defined.

The geometrical worm algorithm~\cite{bosons,directed} has proven to be very efficient for the
study of the bosonic Hubbard model when formulated in terms of currents on the links of the space-time
lattice. The estimated $z_{MC}$ is very close to zero~\cite{directed}. 
It is therefore natural to generalize this algorithm to classical models defined in
terms of bond-variables. This is done in section~\ref{sec:dworms}, where a very efficient 
algorithm formulated directly on the {\it dual} lattice is developed.
A related dual algorithm was initially described by
Prokof'ev and Svistunov~\cite{ProkofevClassical}.
Our algorithm can 
be generalized to Potts, clock and other discrete lattice models of which the Ising model
is the simplest example. The dual algorithm also allows for a very simple and elegant way
of calculating the domain wall free energy directly. In section~\ref{sec:dworms} we outline
how this is done. Regrettably, only in two dimensions is it possible to define such
a dual algorithm. In section~\ref{sec:dworms} we also present a directed version of
this algorithm where the probability for the worms to erase themselves is minimized.
However, before presenting the dual algorithm it is instructive to consider
geometrical worm algorithms defined on the {\it direct} lattice. Such an algorithm
is described in section~\ref{sec:linworms}, in both directed and undirected versions. This algorithm
is applicable in any dimension but is of less interest due to its poor efficiency.
However, from an algorithmic perspective this algorithm is interesting in its own right and it
leads naturally to the definition of the dual algorithm. Finally, in section~\ref{sec:discussion}
we conclude with a number of observations concerning the properties of the algorithms.

\section{\label{sec:linworms}Geometrical Worm Algorithms on the Direct Lattice}
The first algorithm we present we shall refer to as the linear worm algorithm.
This algorithm is closely related to the geometrical worm
algorithms~\cite{bosons,directed}, however, the worms do not form closed loops,
instead they form linear strings (a worm) of flipped spins. A major advantage of
this algorithm is that we can select the length of the
linear worm or even the entire distribution of worm lengths. This could be advantageous
for the study of frustrated or disordered models where other cluster algorithms fail
due to the fact that too ``big" or too ``small" clusters are being generated,
leading to a significant loss of efficiency. 

\subsection{Algorithm A (Linear Worms)}
We begin with a number of useful definitions:
In order to define a working algorithm we consider two configurations of the
spins, $\mu$ and $\nu$ related by the introduction of a worm.
Let $\mu$ denote the configurations of the spins without the worm,
$w$, and $\nu$ the configuration of the spins with the worm. Furthermore, let
$s_1\ldots s_W$ be the sites (or spins) visited by the worm (of length $W$) and let $E_i$ denote the
energy require to flip the spin at site $s_i$ from its position in the
configuration $\mu$, with $-E_i$ the energy required to flip spin $i$
from its position in $\nu$ to that in $\mu$. Note that $E_i$ is defined relative to the spin configuration
$\mu$. In the following we shall make extensive use of the activation or local weight for overturning
a spin, $A^{E_i}_{s_i}$. More precisely, $A^{E_i}_{s_i}$ denotes the
weight for flipping $s_i$ from its position in $\mu$ to that in $\nu$ and $A^{-E_i}_{s_i}$
the weight for going in the opposite direction. As we shall see, $A^{E_i}_{s_i}$ is not uniquely
determined. Here, we shall use 
$A^{E_i}_{s_i}=\min(1,\exp(-\Delta E_i/k_BT))$ although other choices would be equally suitable.
When the worm is moving through the lattice
it will move from the current site $s_i$ to a set of neighboring sites \mathversion{bold}$\sigma$\mathversion{normal} and it
becomes necessary to define the normalization $N_{s_i}=\sum_\sigma A^{E_{i+\sigma}}_{s_i+\sigma}$.
This normalization is used for choosing the next neighbor to visit among the set \mathversion{bold}$\sigma$\mathversion{normal} of neighbors.

If one considers the proofs for the geometrical worm algorithms~\cite{bosons,directed}, it is not
difficult to see that a generalization to classical statistical models defined on the direct
lattice will depend crucially on the fact that the normalization $N_{s_i}$ does {\it not} depend
on the position (up or down) of the spin, $s_i$, at the site itself. For the Ising model, defined
in Eq.~(\ref{eq:Ising}), only nearest neighbor interactions are taken into account and we can
satisfy this requirement simply by not allowing the worm to move to any of the 4 nearest neighbor
sites to the current site $s_i$. In principle, one can consider moving the worm to {\it any} other
sites in the lattice, an aspect of this algorithm of considerable intest.
For the case of longer range interactions we would have to restrict the worm to move only to
sites that the current spin does {\it not} interact with. 
Note that, since we allow sites on the same sub-lattice to be visited then
the $A^{E_i}_{s_i}$ will depend on the {\it order} that we visit the sites since sites neighboring
$s_i$ could have been visited previously by the worm.
As an example of a simple choice for the neighbors we show in Fig~\ref{fig:neighborB} an example
where the worm can move to 4 neighbors around the current site, excluding the 4 nearest neighbors to the
site. 
When defining the set of neighbors \mathversion{bold}$\sigma$\mathversion{normal} for the worm to move
to from $s_i$ one also has to satisfy the trivial property that if $s_{i+1}$ is a
neighbor of $s_i$ then $s_i$ should also be a neighbor of $s_{i+1}$.

We now propose one possible implementation of the geometrical worm algorithm for the Ising model.
A set of suitable neighbors to $s_i$ is defined as outlined above and at the beginning
of the construction of the worm we {\it draw} its length, $W$ from
a normalized distribution. 
We refer to this algorithm as linear worms (A). 
Below we outline the algorithm in pseudo-code.
\begin{enumerate}
\item Choose a random starting site,
$s_{i=1}$ and with normalized probability a length $W$.\label{restartA}
\item Calculate the probability for flipping the spin
 	on that site $A^{E_1}_{s_1}$ with $A^{E_i}_{s_i}=\min(1,\exp(-\Delta E_{s_i}/k_BT))$.
\item Flip the selected spin, $s_i$\label{loopA}.
\item For each of the $k$ neighbors in the set of neighbors \mathversion{bold}$\sigma$\mathversion{normal},
 calculate the weight for
 flipping, $A^{E_{i+\sigma}}_{s_i+\sigma}=\min(1,\exp(-\Delta E_{s_i+\sigma}/k_BT))$.
 Calculate the normalization $N_{s_i}=\sum_\sigma A^{E_{i+\sigma}}_{s_i+\sigma}$
 and the probabilities $p^\sigma_{s_i}=A^{E_i}_{s_i+\sigma}/N_{s_i}$.
\item According to the probabilities $p^\sigma_{s_i}$ select a new site
$s_{i+1}$ among the neighbors to go to and increase $i$ by one, $i\to i+1$.
\item If $i<W$ go to \ref{loopA}.
\item Calculate $A^{-E_{W}}_{s_W}$ and $\bar N_{s_W}$ after the worm is constructed and
with probability
$P_e(w)=1-\min(1,A^{E_1}_{s_1}N_{s_1}/(A^{-E_W}_{s_W}\bar N_{s_W}))$ erase the constructed
worm. Here $A^{E_1}_{s_1}$ and $N_{s_1}$ are calculated {\it before} the worm is
constructed.
\item Goto \ref{restartA}.
\end{enumerate}     

Note that if we decide to construct a worm of length $W=1$ at a given site $s_1$ then
the above algorithm corresponds to attempting a Metropolis spin-flip at that site. \\
See appendix~\ref{app:a} for a proof of the above algorithm.

\subsection{Algorithm B (Directed Linear Worms)}
It is an obvious advantage to have control over the
distribution of the length of the worms. However, if we choose the length
of the worm at the start of the construction of the worm as in algorithm A then we allow
for the worm to backtrack, thereby erasing itself. We can try to eliminate or rather minimize the probability for the
worm to do back-tracking by constructing a directed algorithm. We closely follow 
the method outlined in reference~\onlinecite{directed} in order to construct a
directed algorithm. See also reference~\onlinecite{Syljy} for previous work on directed
algorithms. Let $\sigma_m$ and $\sigma_n$ be among the neighbors, ${\bf\sigma}$, of the site $s_i$.
The above proof of algorithm A does not depend directly on the
definition of the probabilities $p^{\sigma_m}_{s_i}$ and $p^{\sigma_n}_{s_i}$, but
only on their ratio, since they have to satisfy the following relation:
\begin{equation}
\frac{p^{\sigma_m}_{s_i}}{p^{\sigma_n}_{s_i}}=
\frac{A_{s_i+\sigma_m}/N_{s_i}}
{A_{s_i+\sigma_n}/N_{s_i}}=
\frac{A_{s_i+\sigma_m}}
{A_{s_i+\sigma_n}}.
\end{equation}
This leaves us considerable freedom since we can define
conditional probabilities, $p_{s_i}(m|n)$, corresponding to the
probability to continue in the direction $\sigma_m$ at the site $s_i$
if we are coming from $\sigma_n$. At a given site we then only need to satisfy
\begin{equation}
\frac{p_{s_i}(m|n)}{p_{s_i}(n|m)}=\frac{A_{s_i+\sigma_m}}
{A_{s_i+\sigma_n}}.
\label{eq:balance}
\end{equation}
If we have $k$ neighbors this defines a $k\times k$ matrix $P$ where the
diagonal elements corresponds to the back-tracking probabilities, the probabilities for the worm
to erase itself.
Previously, we had effectively been using $p_{s_i}(m|n)$'s which were independent of
the incoming direction, $\sigma_n$, since we simply had
$p_{s_i}(m|n)=A_{s_i+\sigma_m}/N_{s_i}$, corresponding to a matrix $P$
with identical columns.
However, the above condition of balance, Eq.~(\ref{eq:balance}) leaves sufficient
room for choosing very small back-tracking probabilities and in most
situations the corresponding diagonal elements of $P$ can be chosen to be zero. If
we impose the constraint that the sum of the diagonal elements of $P$
should be minimal the problem of finding an optimal matrix $P$ can be
formulated as a standard linear programming problem to which
conventional techniques can be applied~\cite{directed}.

\begin{figure}
\begin{center}
\includegraphics[clip,width=8cm]{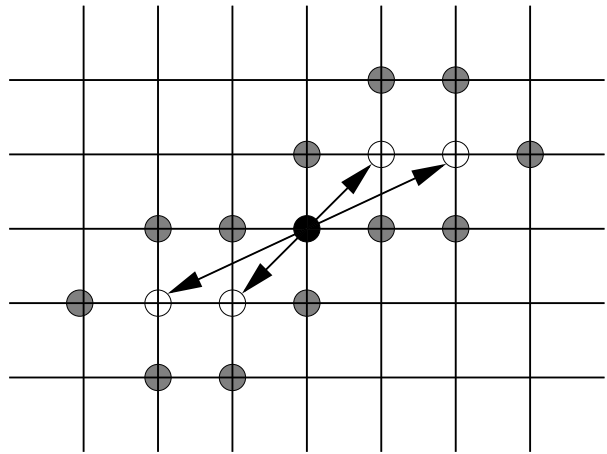}
\caption{An example of the configuration of the neighbors for algorithm A,B. The current site, $s_i$, is
indicated by $\bullet$, the neighbors by $\circ$. The activation (weight) at the
current site will depend on the spins at the nearest neighbor sites (shaded
circles).\label{fig:neighborB}}
\end{center}
\end{figure}

At a given site $s_i$ we choose 4 neighbors, as indicated in
Fig.~\ref{fig:neighborB} by the open circles $\circ$. The activation 
at a given neighbor will depend on its 4 nearest neighbors shown as the
shaded circles in Fig.~\ref{fig:neighborB}. 
Technically, each $p_{s_i}(m|n)$ now depends on the position of all the
16 surrounding spins.
and hence there are in principle $2^{16}$
possible matrices $P$ at each site. It is therefore not feasible to choose too large a set
of neighbors when directing the algorithm.
In the absence of disorder, the $2^{16}$ matrices can be tabulated and minimized at the beginning of
the simulation and for the isotropic ferromagnetic model at hand it is easy to see that at most
625 {\it different} matrices occur.

We can now use the above matrices $P$ for an algorithm that performs
{\it directed} linear worms of length varying between 1 and $W$.
We again assume that a set \mathversion{bold}$\sigma$\mathversion{normal} of $k$ neighbors
has been chosen.

\begin{enumerate}
\item Choose a random starting site,
$s_{i=1}$ and with normalized probability a length $W$.\label{restartB}
\item Calculate the probability for flipping the spin
 	on that site $A^{E_1}_{s_1}$ with $A^{E_i}_{s_i}=\min(1,\exp(-\Delta E_{s_i}/k_BT))$.
\item Flip the selected spin, $s_i$\label{loopB}.
\item {\bf If} $s_i=s_1$ {\bf then} for each of the $k$ neighbors 
in the set \mathversion{bold}$\sigma$\mathversion{normal}, calculate the probability for
 flipping, $A^{E_{i+\sigma}}_{s_i+\sigma}=\min(1,\exp(-\Delta E_{s_i+\sigma}/k_BT))$.
 Calculate the normalization $N_{s_i}=\sum_\sigma A^{E_{i+\sigma}}_{s_i+\sigma}$
 and the probabilities $p^\sigma_{s_i}=A^{E_i}_{s_i+\sigma}/N_{s_i}$.
 {\bf Else:} Depending on the configuration of the nearest neighbors of
 the $k$ neighbors select the correct (minimized) matrix $P$ and if the
 worm arrived from direction $\sigma_n$ set $p^\sigma_{s_i}$ equal to
 the n'th column of $P$.
\item According to the probabilities $p^\sigma_{s_i}$ select a new site
$s_{i+1}$ to go to and increase $i$ by one, $i\to i+1$.
\item If $i<W$ go to \ref{loopB}.
\item Calculate $A^{-E_{W}}_{s_W}$ and $\bar N_{s_W}$ after the worm is constructed and
with probability
$P_e(w)=1-\min(1,A^{E_1}_{s_1}N_{s_1}/(A^{-E_W}_{s_W}\bar N_{s_W}))$ erase the constructed
worm. Here $A^{E_1}_{s_1}$ and $N_{s_1}$ are calculated {\it before} the worm is
constructed.
\item Goto \ref{restartB}.
\end{enumerate}     
The proof of the above algorithm can be found in appendix~\ref{app:b}.

\subsection{Performance: Algorithms A,B}
\begin{figure}
\begin{center}
\includegraphics[clip,width=8cm]{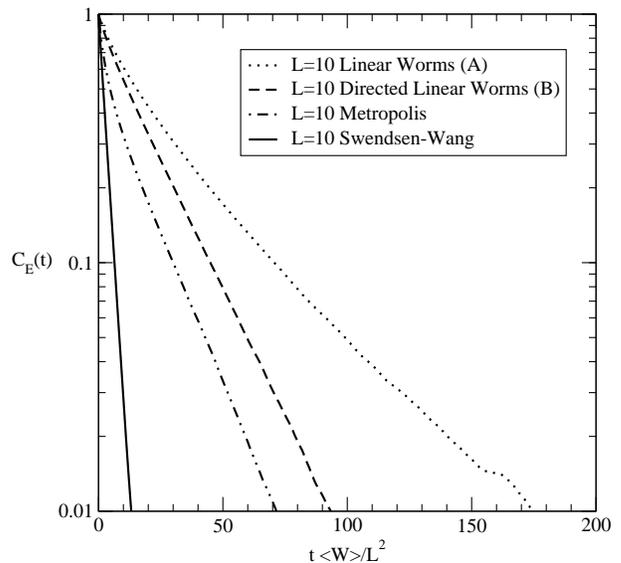}
\caption{Autocorrelation functions for the energy $C_E(t)$ as a function of Monte Carlo time, calculated
at $T_c$. 
Shown are results for $L=10$ for the Metropolis, Swendsen-Wang and for the 
linear worm (A) and directed linear worm (B) algorithms (with uniform worm length).
Note that in order to compare the 4 algorithms the time axis has been scaled so that in all cases
1 time step corresponds to an attempted update of all the variables. 
The autocorrelation functions shown were calculated averaging over 20 million worms
(MCS for the metropolis algorithm).}
\label{fig:PerfAB}
\end{center}
\end{figure}

In order to test the performance of algorithms A and B we consider the autocorrelation
function
$C_{\cal O}(t)$
of an observable ${\cal O}$.
This function is defined
in the standard way:
\begin{equation}
C_{\cal O}(t)=\frac{\langle {\cal
O}(t){\cal O}(0)\rangle - \langle {\cal O} \rangle^2}{\langle {\cal O}^2
{\rangle - \langle {\cal O} \rangle}^2}
\end{equation}
For a reliable estimate of $C_{\cal O}(t)$ we typically generate 20 million worms. In Fig.~\ref{fig:PerfAB}
we show results for the autocorrelation function for the energy $C_E(t)$ for the linear worm
algorithms A,B for a system of size $L=10$. In both cases the worm length $W$ was chosen
from a uniform distribution. This is compared to $C_E(t)$ for the usual single flip
Metropolis algorithm and the Swendsen-Wang algorithm. For the latter two algorithms the Monte Carlo
time is usually measured in terms of Monte Carlo Sweeps (MCS) where an attempt to update {\it all}
the $L^2$ spins in the lattice has been made. However, for algorithms A and B a worm will on average
only attempt to update $\langle W\rangle$ of the $L^2$ spins. Hence, if time is measured in terms of generated
worms for the worm algorithms it should be rescaled by a factor of $\langle W\rangle/L^2$
for a fair comparison to be made
with algorithms where Monte Carlo time is measured in MCS. This has been done in Fig.~\ref{fig:PerfAB}.
From the results in Fig.~\ref{fig:PerfAB} it is clear that the efficiency of algorithm A and B is
fairly poor and worse than the much simpler Metropolis algorithm. We have checked that this remains
true for significantly larger lattices sizes.

The main cause of this poor behavior
is the necessity to include a rejection probability $P_e$. If a worm on average attempts to update
$\langle W\rangle$ spins, we need on average to generate $L^2/\langle W\rangle$ worms to complete a MCS.
As a rough estimate, let us consider that the worm is accepted with the average probability $p$ 
and that only a fraction $x$ of the $\langle W\rangle$ attempted spin flips result in an updated spin (a spin can 
be visited several times). It then follows that on average $pxL^2$ are actually flipped in a MCS
with the linear worm algorithm. If the average probability for flipping a spin with the Metropolis
algorithm is $q$ and if all the spins selected in a MCS are different, the corresponding number
for the Metropolis algorithm is $qL^2$, presumably of the same order as $pxL^2$ and likely larger. 
The effect of the directed linear worm algorithm is to maximize $x$ as much as possible and 
it therefore
seems unlikely that the linear worm algorithm in its present form can perform better than the Metropolis algorithm 
unless $p$ also is maximized.

It would be very interesting if algorithms A and B could be modified so that $p\equiv 1$ ($P_e\equiv 0$). 
We have so far been unable to do so. Such a modified algorithm would be significantly more efficient than
the Metropolis algorithm (but presumably
less efficient than the Swendsen-Wang algorithm). It would be much more versatile and could
be of significant interest for frustrated or disordered systems since it would not require the construction
of clusters but only of the much simpler worms, the length of which can be chosen. 

Even though it is quite interesting to be able to choose the distribution of the worm
lengths at the outset, the question arises which distribution of worm lengths will give the most
optimal algorithm. In the present work we have chosen a distribution of worm lengths that is uniform between
1 and $2L$, but on general grounds we expect a power-law form for this distribution to be more optimal at the critical point.
For the study of Bose-Hubbard models using geometrical worm algorithms it was noted~\cite{directed} that the distribution of
the size of 
the worms follow a power-law with an exponent of approximately 1.37 at the critical point.  Hence, in the present case it
would appear likely that an optimal power-law distribution of the worm lengths at $T_c$ can be found, defining a new
``dynamical" exponent.
The idea of choosing the distribution of the worm size to optimize the algorithm
resembles previous work by Barkema and Newman~\cite{barkema93,newman96}.

In closing this section, we note that it is quite straightforward to define an algorithm on the direct lattice
where the worms form closed loops (rings). In this case the length of the worm is determined
by the size of the loop. We have tested such algorithms but their performance
is even worse than algorithms A and B since the constraint that the initial spin has to be 
revisited makes the length
of the worms in some cases diverge, or for other variants of such an algorithm, go to zero.

\section{\label{sec:dworms}Dual Algorithm}
It is clear that one problem with both of the above algorithms is the fact that
the spin on the initial site is treated differently than the remaining spins. This
is because the geometrical worm algorithms are more suitable for an implementation
directly on the dual model. Hence, we will now try to describe an algorithm that
moves a worm along the {\it dual} lattice by updating bond-variables on the direct lattice.

We begin with some definitions analogous to the treatment of Kadanoff~\cite{kadanoff}.
On the direct lattice we define, at each site $(j,k)$ an integer variable $s_{j,k}=\pm 1$.
Here $j$ describes the index in the x-direction and $k$ the index in the y-direction.
Then we can define the following bond-variables:
\begin{eqnarray}
b^x(j,k)=s_{j+1,k}s_{j,k}\\
b^y(j,k)=s_{j,k+1}s_{j,k}.
\end{eqnarray}
The Ising model has $L^2$ variables where as we see that we have $2L^2$ bond-variables.
However, it is easy to see that the bond-variables satisfy a divergence free constraint
at each site of the dual lattice:
\begin{equation}
b^x(j,k)+b^y(j+1,k)-b^x(j,k+1)-b^y(j,k) ({\rm mod} 4) = 0
\label{eq:divfree}
\end{equation}
giving us $L^2$ constraints. However, if we define the model on a torus the constraint
on each dual lattice site is equal to the sum over the constraints on all the other dual
sites. Hence, in this case we obtain only $L^2-1$ independent constraints.
In addition we also have to satisfy the boundary conditions,
$s_{L+1,k}\equiv s_{1,k},\ \ s_{j,L+1}\equiv s_{j,1}$.
This implies that for an $L\times L$ lattice with {\it even} $L$:
\begin{eqnarray}
\sum_{j=1}^{L}b^x(j,k)-L \ ({\rm mod} 4) =0\ \forall k\\
\sum_{k=1}^{L}b^y(j,k)-L \ ({\rm mod} 4) =0\ \forall j.
\label{eq:bcconstraints}
\end{eqnarray}
These constraints are not independent of the previously defined constraints Eq.~(\ref{eq:divfree}).
In fact, it is easy to see that if the boundary constraints Eq.~(\ref{eq:bcconstraints}) are applied
at just one row and one column then the divergence free constraints Eq.~(\ref{eq:divfree}) will enforce
the boundary constraints at the remaining rows and columns. Hence, these constraints only give us
2 more independent constraints, in total $L^2+1$ constraints. The model, written in terms of the
bond-variables, therefore has $L^2-1$ free variables and correspondingly half the number of degrees of
freedom compared to the formulation in terms of the spin-variables. This is a natural consequence of the fact
that the bond-variable model does not distinguish between a state and the same state with all the spins
reversed.

The partition function for the Ising model on a $L\times L$ torus can now be written in the following manner:
\begin{eqnarray}
&Z={\rm Tr}'_{b^x}{\rm Tr}'_{b^y} \prod_{j=1}^L\prod_{k=1}^L\nonumber\\
&\times \exp\left\{K^x_{j,k} b^x(j,k)
                +K^y_{j,k} b^y(j,k) \right\}.
\label{eq:partfunction}
\end{eqnarray}
Here Tr$'$ denotes the trace over bond-variables satisfying the above constraints
and $K^x_{j,k}=J^x_{j,k}/k_B T,\ K^y_{j,k}=J^y_{j,k}/k_B T$.

\begin{figure}
\begin{center}
\includegraphics[clip,width=8cm]{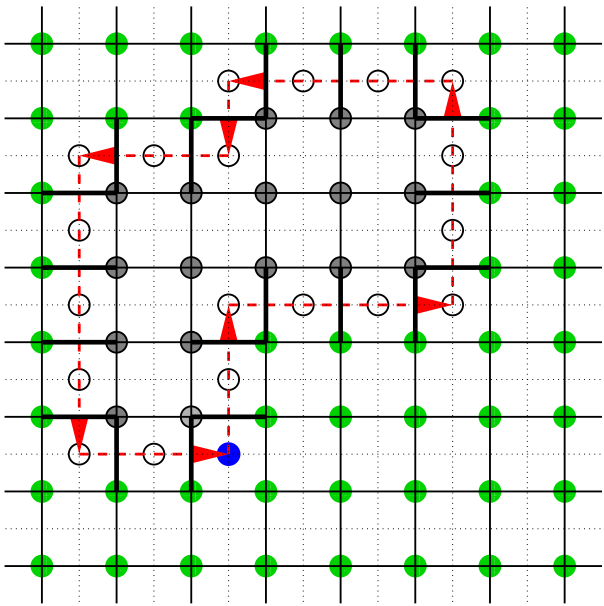}
\caption{A worm moving through the dual lattice. The direct lattice is indicated
by solid lines and the spins on the direct lattice by solid circles. The dual lattice
is indicated by dashed lines and the sites on the dual lattice by open circles. As the
worm moves through the lattice the bond-variables are flipped as indicated
by the thick bonds in the figure.}
\label{fig:dualworm}
\end{center}
\end{figure}

\begin{figure}
\begin{center}
\includegraphics[clip,width=8cm]{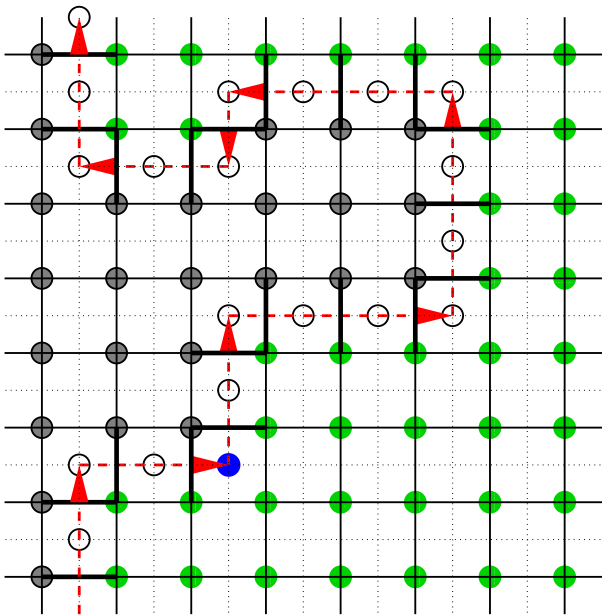}
\caption{An illegal worm move. The worm winds around the lattice in 
the vertical direction and correspondingly {\it all} rows have an {\it odd}
number of flipped bond-variables $b$. The direct lattice is indicated
by solid lines and the spins on the direct lattice by solid circles. The dual lattice
is indicated by dashed lines and the sites on the dual lattice by open circles. As the
worm moves through the lattice the bond-variables are flipped as indicated
by the thick bonds in the figure.}
\label{fig:dualillegalmove}
\end{center}
\end{figure}

\subsection{Algorithm C (Dual Worm Algorithm)}
We now turn to a discussion of the dual algorithm. From the above description in
terms of bond-variables it is now quite easy to define a geometrical worm algorithm on
the dual lattice closely
following previous work on such algorithms~\cite{bosons,directed}.
We denote the i'th site on the dual lattice that the worm visits as $d_i$. A ``worm" is constructed
by going through a sequence of neighboring sites on the dual lattice by each time choosing a direction
$\sigma$ to follow according to an appropriately determined probability. When the worm moves
from $d_i$ to $d_{i+1}$ the corresponding bond-variable $b_i$ that the worm crosses is flipped.
The bond-variables
can take on only 2 values $\pm 1$. Hence, the associated energy cost for flipping $b_i$ is given by
$\Delta E= 2J^\alpha_{j,k}b^\alpha(j,k), \ \alpha=x,y$.
We can now define weights for each direction $\sigma$,  $A^{E_\sigma}=\min(1,\exp(-\Delta E^{\sigma}/k_BT))$
used for determining the correct probability for choosing a new site.
This is not the only choice for the weights. Other equivalent choices should work equally well.
The bond-variables are updated during the construction of the worm and the worm is finished
when the starting site on the dual site is reached again and the worm forms a closed loop.
The dual worm algorithm can then be summarized
using
the following pseudo-code:
\begin{enumerate}
\item  Choose a random initial site 
$d_1$ on the dual lattice.\label{dualstart}
\item For each of the directions
$\sigma=\pm x,\pm y$ calculate the weights
$A^{E_\sigma}$ associated with flipping the bond-variable perpendicular to that direction,
$A^{E_\sigma}=\min(1,\exp(-\Delta E^{\sigma}/k_BT))$.\label{Uloop}
\item Calculate the normalization $N_{d_i}=\sum_\sigma A^\sigma$
and the associated probabilities
$p^\sigma_{d_i}={A^{E_\sigma}}/{N_{d_i}}.$
\item According to the probabilities, $p^\sigma_{d_i}$,
choose a direction $\sigma$.
\item Update the bond-variable $b^\sigma_{i}$ for the direction chosen
and move the worm to the new dual lattice site $d_{i+1}$.
\item If $d_i\ne d_1$ goto \ref{Uloop}.
\item Calculate the normalizations $\bar N_{d_1}$ and $N_{d_1}$
of the initial site, $s_1$, with and without the worm present. If the
worm is ``legal", i.e. with {\it even} winding number in both the x- and the y-direction
($O^x_w,\ O^y_w$ both 0 - see definition below),
then erase the worm with probability $P_e=1-\min(1,N_{d_1}/{\bar N_{d_1}})$. If the worm
is ``illegal", that is if either $O^x_w$ or $O^y_w$ calculated when the worm
has closed equal 1, then always erase it. Goto \ref{dualstart}
\end{enumerate}

Following the above discussion of the boundary constraints it is easy determine if
the winding number in the x- or y-direction is odd by simply choosing one row $k_0$
and one column $j_0$ and calculating the number of frustrated, $b=-1$, bond-variables:
\begin{eqnarray}
O^x&=&\sum_j \frac{1-b^x(j,k_0)}{2}\ ({\rm mod} 2)\nonumber\\
O^y&=&\sum_k \frac{1-b^y(j_0,k)}{2}\ ({\rm mod} 2),
\label{eq:windings}
\end{eqnarray}
since a worm with an odd winding number in the y-direction will result in $O^x$ being
1 independent of $k_0$, or equivalently for the x-direction and $O^y$. See Figure~\ref{fig:dualillegalmove}.
$O^x$ and $O^y$ then takes on the values 1 or 0 depending on whether the corresponding
winding numbers are odd or even.

Several points are noteworthy about this algorithm. To a great extent it is simply
the dual version of the Wolff algorithm~\cite{Wolff}. Each worm encloses a cluster
of spins on the direct lattice (not necessarily a simply connected cluster). Flipping
the bond-variables in the worm effectively flips the spin in the cluster. This correspondence
is particular to two dimensions which is the only dimension where the present dual worm algorithm can be defined.
The intuitive argument for this is that the bond-variables updated by the worm have to enclose a finite
volume of spins on the direct lattice, this is only possible in two dimensions. In dimensions higher
than two the one dimensional worms are not capable of enclosing a finite volume. Mathematically, it can
be seen that the divergence less constraint, Eq.~(\ref{eq:divfree}), cannot be satisfied under worm moves.
Under most conditions the rejection probability, $P_e$, is very small, however, for small
lattice sizes the probability for generating a worm with an {\it odd} winding number
in either the x- or y-direction can be non-negligible.
It is important to note that since the bond-variables are updated during the construction of
the worm the generated
configurations are not valid and the above constraints are {\it not} satisfied.
However, once the construction of the worm is
finished and the path of the worm closed, the divergenceless constraint is
satisfied. Rejecting worms with odd winding numbers then assures that the resulting
configurations are valid. Also, when the worm moves through the lattice it may pass
many times through the same link and cross itself before it reaches the
initial site where the construction terminates. 
Finally, at each step $i$ in the construction of the worm it is
likely that the worm at the site $d_i$ will partially ``erase" itself by
choosing to go back to the site $d_{i-1}$ visited immediately before, 
thereby ``bouncing" off the site $d_i$.\\

\noindent\underline{\bf Proof of Algorithm C (Dual Worm Algorithm)}\\

Now we turn to the proof of detailed balance for the
algorithm.
As before, let $\mu$ denote the configuration of the bond-variables without the worm,
$w$, and $\nu$ the configuration with the worm. 
Let us consider the case where the
worm, $w$, visits the sites $\{d_1\ldots d_W\}$ on the dual lattice.
Here $d_1$ is the initial
site. The worm then goes through the corresponding bond-variables
$\{b_1\ldots b_W\}$.
Note that, $d_W$ is the last site visited before the worm reaches
$d_1$. 
Furthermore, let $E_i$ denote the
energy required to flip the bond-variable $b_i$ from its position in the
configuration $\mu$, with $-E_i$ the energy required to flip $b_i$
from its position in $\nu$ to that in $\mu$. 
The total probability for constructing the worm $w$ is then given by:
\begin{eqnarray}
&&P(w;d_1\to d_W) = \nonumber\\
&&P(d_1)(1-P_e(w))P({\rm legal}|w)\prod_{i=1}^W \frac{A^{E_i}}{{N_{d_i}}}.
\end{eqnarray}
The index $\sigma$ denotes the direction needed to go from
$d_i$ to $d_{i+1}$, ($\pm x, \pm y$), crossing the bond-variable $b_i$.
$P(d_1)$ is the probability for choosing site $d_1$ as the starting
point and $P_e(w)$ is the probability for erasing a ``legal" worm after
construction. Finally, $P({\rm legal}|w)$ is the conditional probability that
the constructed worm $w$ is ``legal", with {\it even} winding numbers in both
x- and y-directions.
If the worm $w$ is legal and has been accepted we have to consider the
probability for reversing the move. That is, we consider the probability for constructing an anti-worm
$\bar w$ annihilating the worm $w$. 
Note that, in this case the sites are visited in the opposite order,
$\bar d_1 = d_1, \bar d_2 = d_{W},
\ldots, \bar d_W = d_2$, in general $\bar d_i=d_{W-i+2}\ (i\neq 1)$. 
In the same manner the bond-variables are also visited in the reverse
order
$\bar b_1 = b_W, \bar b_2 = b_{W-1},
\ldots, \bar b_W = b_1$, in general $\bar b_i=d_{W-i+1}$. 
\begin{figure}
\begin{center}
\includegraphics[clip,width=8cm]{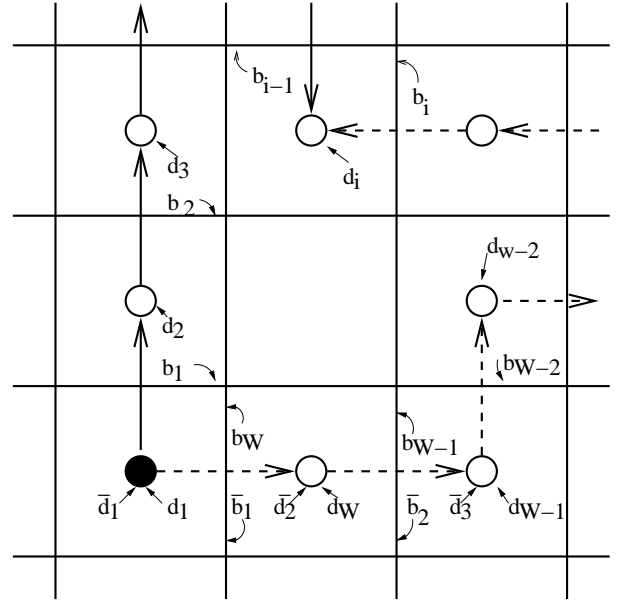}
\caption{The notation used for the proof of the dual worm algorithm.
The configuration shown corresponds to a
worm (solid line) partly erased by its corresponding anti-worm (dashed line).}
\label{fig:dualnotation}
\end{center}
\end{figure}
Consequently, when describing $\bar w$, we directly use $d,b$ instead of $\bar d,\bar b$.
The notation is illustrated in Fig.~\ref{fig:dualnotation}.
We therefore have:
\begin{eqnarray}
P(\bar w;\bar d_1\to \bar d_W)&=&
P(d_1)(1-P_e({\bar w}))P({\rm legal}|{\bar w})\nonumber\\
&\times&\frac{A^{-E_{W}}}{{\bar N}_{d_1}}\prod_{i=W}^2\frac{A^{-E_{i-1}}}{{\bar N}_{d_i}}.
\end{eqnarray}
Please note that the index $i$ runs over $W,\ldots, 2$.
Let us now consider the case where both of the worms $w$ and $\bar w$ have reached the site
$d_i$ {\it different} from the starting site $d_1$. See Fig.~\ref{fig:dualnotation}.
Since we are updating the link variables during
the construction of the worm we immediately see that $N_{d_i}=\bar N_{d_i}\ i\neq 1$.
Furthermore, $A^{E_{i}}$ and $A^{-E_{i}}$
only depend on the bond-variable $b_i$,
and we therefore see that:
\begin{equation}
\frac{A^{E_i}}
{A^{-E_i}}=
\exp(-\Delta E_{i}/k_BT),\ i=1\ldots W.
\end{equation}
Hence, since $P(d_1)$ is uniform through out the dual lattice, and since
$w$ and $\bar w$ must wind the lattice
in precisely the same manner resulting in 
$P({\rm legal}|w)=P({\rm legal}|{\bar w})$, 
we find:
\begin{equation}
\frac{P(w)}
{P({\bar w})}
=\frac{1-P_e({w})}{1-P_e({\bar w})}
\frac{\bar N_{d_1}}{N_{d_1}}
\exp(-\Delta E_{\rm Tot}/k_BT),
\end{equation}
where $\Delta E_{\rm Tot}$ is the total energy difference between a
configuration with and without the worm $w$ present. With our definition
of $P_e$ we see that $(1-P_e(w))/(1-P_e(\bar w))=N_{d_1}/{\bar N_{d_1}}.$
Hence, with this choice of $P_e$ we satisfy detailed balance since:
\begin{equation}
\frac{P_w}{P_{\bar w}}=
\exp(-\Delta E_{\rm Tot}/k_BT).
\label{eq:partbal}
\end{equation}
As was the case for the previous algorithms there are 2 ways of introducing a worm
that will take us from $\mu\to\nu$, and equivalently 2 ways of introducing $\bar w$.
(Strictly speaking, each of the two ways actually represents an infinite sum as discussed
under the proof of algorithm A).
In both cases Eq.~(\ref{eq:partbal}) holds and 
it follows that $P(\mu\to\nu)=P(\nu\to\mu)\exp(-\Delta E_{\rm Tot}/k_BT)$ showing that
detailed balance is satisfied.
Ergodicity is simply proven as the worm can
perform local loops  and wind around the
lattice in any direction. $\blacksquare$

{\it Illegal worms:} One should be cautious when treating  ``illegal" moves such as
the worms with odd winding numbers encountered above. Suppose that our Monte Carlo algorithm
has generated a number of configurations $C_1,C_2,C_2,C_3,\ldots,C_{i-1},C_{i}$ where some
configurations are repeated since the worm moved has been rejected. From the configuration
$C_i$ we now construct a worm that turns out to be ``illegal" with odd winding numbers. From the above
proof we then see that we {\it have} to repeat $C_i$ in the sequence of configurations since {\it not}
repeating the configuration would lead to a total acceptance probability different from
$(1-P_e(w))P({\rm legal}|w)$. For instance, imagine we kept generating worms till we found a ``legal"
one. This would make the total acceptance ratio a sum over the number of times we try which would
be incorrect.

\subsection{Algorithm D (Dual Directed Worm Algorithm)}
Following the discussion of algorithm B it is straight forward to develop a directed version of algorithm C.
As was
the case for algorithm B we define conditional probabilities $p_{d_i}(m|n)$, corresponding to the probability
for continuing in the direction $\sigma_m$ if the worm has come from direction $\sigma_n$. Since we in
algorithm C, at each
site on the dual lattice $d_i$, can go in four directions this leads us to define a $4\times 4$ matrix P
at each site on the dual lattice. However, compared to algorithm B, the number of different matrices
at a given dual site is now much smaller since the weights $A^{E_i}$ {\it only} depend on the associated
bond-variable. In fact there are now only 16 possible matrices at a given site and these can easily be
optimized and tabulated at the outset of the calculation. Even in the presence of disorder, a large number
of them can be tabulated. This directed dual worm algorithm is now identical to algorithm C except for
the fact that if $d_i$ is different from $d_1$, the $p_{d_i}^\sigma$ are selected from these optimized
matrices (the same way at it was done for algorithm B). Due to the simplicity of this modification
and the similarity with algorithm B,
we do not present pseudo-code nor a proof for this directed algorithm. It is easy to see that rejection
probability remains identical to the one used in algorithm C, $P_e(w)=1-\min(1,N_{d_1}/\bar N_{d_1})$.

\subsection{Performance Algorithm C,D}
In order to compare the two algorithms C and D defined on the dual lattice we have again
calculated the autocorrelation function $C_E(t)$.
\begin{figure}
\includegraphics[clip,width=8cm]{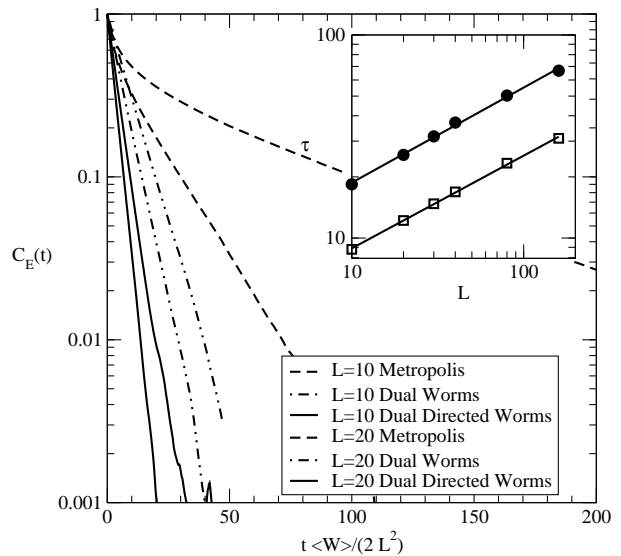}
\caption{Autocorrelation functions for the energy $C_E(t)$ as a function of Monte Carlo time, calculated
at $T_c$. 
Shown are results for $L=10,20$ for the Metropolis, dual worm and dual directed worm algorithms.
Note that, in order to compare the 3 algorithms the time axis has been scaled so that in all cases
1 time step corresponds to an attempted update of all the variables. The inset shows the scaling
of the autocorrelation time $\tau'$ as a function of the system size for the dual worm and dual directed worm
algorithm. The autocorrelation time is here rescaled:
$\tau'=\tau \langle$W$\rangle/(2L^2)$. In both cases we find roughly a power-law form $\tau'\simeq L^{0.46}$ shown as the solid lines
in the inset. The autocorrelation functions shown were calculated averaging over 10-20 Million worms
(MCS for the metropolis algorithm).}
\label{fig:ACDual}
\end{figure}
\begin{table}
\caption{\label{tab:table1}Average worm size defined as the average number of bond-variables attempted to
change during the construction of a any worm, including worms eventually rejected. Also listed is
the estimated autocorrelation time, $\tau$, in units of worms constructed and the rescaled autocorrelation time
$\tau'=\tau \langle$W$\rangle/(2L^2)$ in units of $2L^2$ attempted updates. The autocorrelation time is here
here defined by $C_e(\tau)=0.05$. Results are listed for the dual and dual directed worm algorithms.
}
\begin{ruledtabular}
\begin{tabular}{rrrrrrr}
 &\multicolumn{3}{c}{Dual}&\multicolumn{3}{c}{Dual Directed}\\
L&$\langle$W$\rangle$&$\tau$&$\tau'$&$\langle$W$\rangle$&$\tau$&$\tau'$\\
\hline
10  &   57 &  64.6  & 18.4 &   60 &  29.3 & 8.8\\
20  &  188 &  109.4 & 25.7 &  198 &  49.3 & 12.2\\
30  &  382 &  149.4 & 31.7 &  400 &  66.6 & 14.8\\
40  &  630 &  187.9 & 37.0 &  660 &  81.9 & 16.9\\
80  & 2119 &  303.8 & 50.3 & 2204 & 135.9 & 23.4\\
160 & 7132 &  478.8 & 66.7 & 7408 & 214.3 & 31.0\\
\end{tabular}
\end{ruledtabular}
\end{table}
Our results are shown in Fig.~\ref{fig:ACDual} and  tabulated in Table~\ref{tab:table1}. As on the direct
lattice we use the energy to calculate the autocorrelation functions. For algorithms C and D we have constructed
10 million worms for the smaller lattices and 20 million for the largest lattice (L=160). We compare to the
Metropolis algorithm where each time step corresponds to an attempted update of the $L^2$ spins. Since the
worm algorithm in this case on average only attempts to update $\langle W\rangle$ out of the $2L^2$ bond-variables we
have to rescale the time-axis by a factor of $\langle W\rangle/(2L^2)$ when we compare algorithms C and D 
to the Metropolis algorithm. 
We can therefore define an autocorrelation time $\tau$ in units of worms constructed and a rescaled
autocorrelation time $\tau'=\tau \langle W\rangle/(2L^2)$ in units of $2L^2$ 
attempted updates. In order to simplify
the analysis of the data we define the autocorrelation time as $C_E(\tau)\equiv 0.05$. As can be seen in
Fig.~\ref{fig:ACDual} the dual worm algorithms represents a dramatic improvement over
the Metropolis algorithm. We have not plotted results for the Swendsen-Wang algorithm on the direct lattice since
for $L=10$ they are indistinguishable from the results obtained for the dual directed worm algorithm.
The scaling of $\tau'$ with $L$ is a relatively small power-law $\tau'\sim L^{0.46}$. This exponent is small
but {\it larger} than most values quoted for the Swendsen-Wang and Wolff algorithms. This is perhaps not too
surprising since the worms are allowed to cross themselves thereby erasing previous updates. The difference
in autocorrelation times between algorithm C and its directed version D appears to be a constant
factor. Therefore, the direction of algorithm C does not change the exponent $z_{MC}$. 

For a $10\times 10$ lattice at $T_c$ roughly 85\% of all constructed worms are accepted and about 10\% of the
constructed worms are ``illegal" when simulations are performed using the dual geometric worm algorithm. For 
the directed version of the algorithm the corresponding numbers are 75\% and 20\%. The number of illegal worms
drops rapidly with $L$ and for $L=30$ 90\% of the worms are accepted with 5\% illegal worms for the undirected
algorithm compared to 83\% and 10\% for the directed dual algorithm.

The presented
dual algorithms are quite general since they depend only in a relatively limited way on the underlying
Hamiltonian.  Even though they may not be competitive with the Swendsen-Wang and
Wolff algorithms for the Ising model they may be quite interesting to apply to other two-dimensional models that
can be formulated in terms of bond-variables and where more effective cluster algorithms are difficult to define. 

\subsection{\label{sec:dwall}Domain Wall Free Energy}
A quantity of particular interest in many studies of ordering transitions is the domain wall free energy,
the difference in free energy between configurations of the system with periodic (P) and anti-periodic (AP) boundary
conditions in one direction, $\Delta F=F_{\rm AP}-F_{\rm P}$.
The domain wall free energy can be shown to obey simple scaling relations and is a very efficient tool for
distinguishing between different ordered phases. Unfortunately, it is usually very difficult to calculate free
energies directly in Monte Carlo simulations. A clever trick to do so is the boundary flip MC method,
proposed by Hasenbusch~\cite{Hasenbusch,Hukushima}, where the coupling constants at the boundary are considered 
as dynamical variables, and can be flipped during the course of the MC simulation.
If $P_P(T)$ and $P_AP(T)$ describe the probability to obtain MC configurations with periodic and anti-periodic
boundary conditions and
and $Z_{\rm P}, Z_{\rm AP}$ the associated partition functions, 
then the domain wall free-energy is given by:
\begin{equation}
e^{-\beta(F_{\rm AP}-F_{\rm P})}=\frac{Z_{\rm AP}}{Z_{\rm P}}=\frac{P_{\rm AP}(T)}{P_{\rm P}(T)}.
\end{equation}
Effectively, the boundary flip MC method attempts to sample the ratio $Z_{\rm AP}/Z_{\rm P}$, and this is usually
cumbersome. However, for the dual algorithms it is trivial to obtain this ratio.
It follows from Eq.~(\ref{eq:bcconstraints}), that if we allow the constructed worms to have all different
kinds of winding around the torus (making {\it all} worms legal) then we are sampling a partition function
which is a sum of four terms:
\begin{equation}
Z=Z_{\rm P_x,P_y}+Z_{\rm AP_x,P_y}+Z_{\rm P_x,AP_y}+Z_{\rm AP_x,AP_y}
\end{equation}
Here, $Z_{\rm BC_x,BC_y}$ is the partition function with BC$_x$(BC$_y$) in the x(y)-direction. 
This is so, because we can turn an illegal worm into a legal worm by changing the boundary
condition in the direction(s) where the winding of the worm violates Eq.~(\ref{eq:bcconstraints}). 
During the
MC simulation it is easy to see to which term a given configuration contributes simply by keeping track
of the {\it total} winding number in both the x- and y-direction. See eq.~(\ref{eq:windings}). When this
is even, the boundary condition in the associated direction is periodic, and if it is odd the boundary
condition is anti-periodic. In order to calculate $Z_{\rm AP_x,P_y}/Z_{\rm P_x,P_y}$ we therefore simply
have to count how many times the {\it total} winding number is odd in the x-direction and even in the y-direction
and divide with the number of times it is even in both direction. Due to its simplicity, we expect this approach
to be an extremely efficient way of calculating $\Delta F$. In table~\ref{tab:table2}
we show preliminary results for $\Delta F/kT$ calculated with $10^8$ worms for the two-dimensional
Ising model at $T_c$. For reference we compare to exact results obtained using pfaffians~\cite{pfpure}.
   
We are grateful to Philippe de Forcrand for suggesting the above way of calculating $\Delta F$.
\begin{table}
\caption{\label{tab:table2} The domain wall free-energy $\Delta F/kT$ for several system sizes
calculated at the critical temperature $T_c$ obtained using our MC method and exact results from
Ref.~\onlinecite{pfpure}.
}
\begin{ruledtabular}
\begin{tabular}{rcr}
L& Exact&MC\\
\hline
4 & 0.9658246670    & 0.9654(5)\\
16& 0.9853282229     & 0.985(10) \\
24& 0.9859725679    & 0.985(10)\\
32& 0.9861972559    & 0.986(10)\\
48& 0.9863574947    & 0.986(10)\\
64& 0.9864135295    & 0.984(20)\\
\end{tabular}
\end{ruledtabular}
\end{table}

\section{\label{sec:discussion}Discussion}
We have presented two generalizations of the geometrical worm algorithm applicable
to classical statistical mechanics model. The first algorithm exploits linear worms on the direct
lattice, the other closed loops of worms on the dual lattice. In both cases have we developed directed versions
of the algorithms. While the first algorithm is applicable
in any spatial dimension, it is quite clear that for topological reasons it is only possible to define a worm algorithm
of the proposed kind on the dual lattice in two spatial dimensions.

Due to its poor performance the linear worm algorithm (A) is mainly of interest from the perspective of fostering
the development of new and more advanced algorithms. The linear worm algorithm presents a number of very attractive
features such as the possibility to choose the distribution of the worm lengths and it seems possible to significantly
improve the performance of the algorithm (and most other geometrical worm algorithms) by eliminating the
rejection probability. So far, we have been unable to do so, but this would clearly be very interesting since the 
algorithms then would be quite promising for the study of frustrated or even disordered models where other cluster algorithms fail.

The dual worm algorithm is very efficient and even though the obtained $z_{MC}$ is larger than for the Swendsen-Wang
algorithm it is quite competitive since it requires relatively little overhead and is very simple to implement. 
One of the most interesting features of this algorithm is the topological aspect of the worms, resulting in forbidden
worm moves.  

The
geometrical worm algorithm has in some cases been very successfully applied to the study of models with disorder~\cite{bosons}.
We therefore applied both the linear and dual worm algorithm to the bond disordered Ising model. In both cases do the algorithms
perform worse than the Metropolis algorithm. For the dual worm algorithm this failure is due to the fact that the average
size of worms diverge when glassy phases are approached. The linear worm algorithm would seem more promising for study
of models with disorder since the length of the worms can be controlled from the outset. In fact the performance of the algorithm
is in this case quite comparable to the Metropolis algorithm (which has a diverging $\tau$ close to a glassy phase). One might
have expected the linear worm algorithm to perform significantly {\it better} than the Metropolis algorithm when disorder
is present. We assume that the fact that it is only comparable to the Metropolis algorithm
is because large parts of the spins are not effectively flipped since worms are mostly rejected in those
parts of the lattice. If this is true, a modified version of the linear worm algorithm with $P_e\equiv 0$ or with
a much more ingenious choice of the neighbors would be of considerable
interest since it would hold the potential to sample such disordered models significantly more effectively than the Metropolis
algorithm.

\acknowledgments
This research is supported by NSERC of Canada as well as SHARCNET. PH gratefully acknowledges
SHARCNET for a summer scholarship. FA acknowledges support from the Swiss National Science
Foundation and stimulating discussions with Philippe de Forcrand.

\appendix
\section{Proof of Algorithm A\label{app:a}}
The probability for creating
the worm, $w$, starting from site $s_1$ is given by:
\begin{eqnarray}
P(w;s_1\to s_W)=&&
P(W)P(s_1)
\frac{A^{E_2}_{s_2}}{N_{s_1}}
\frac{A^{E_3}_{s_3}}{N_{s_2}}
\ldots\nonumber\\
&&\times \frac{A^{E_W}_{s_W}}{N_{s_{W-1}}}
(1-P_e(w))
\end{eqnarray}
Here $P(W)$ denotes the probability for choosing a worm length of $W$.
The probability for introducing an anti-worm, $\bar w$, starting at the site
$s_1$, erasing the worm by going in the opposite direction along
$w$ becomes:
\begin{eqnarray}
P(\bar w;s_{W}\to s_1)=&&
P(W)P(s_W)
\frac{A^{-E_{W-1}}_{s_{W-1}}}{\bar N_{s_W}}
\ldots\nonumber\\
&&\times \frac{A^{-E_2}_{s_2}}{\bar N_{s_3}}
\frac{A^{-E_{1}}_{s_{1}}}{\bar N_{s_2}}(1-P_e(\bar w))
\end{eqnarray}
Since neither $N_{s_i}$ nor $\bar N_{s_i}$ depend on the
position of the spin on the site $i$ and since the spins are visited in
the precise opposite order for the anti-worm with respect 
to the worm, we then see that $N_{s_i}=\bar N_{s_i}$ $\forall i\neq 1,W$, and we find:
\begin{equation}
\frac{P(w;s_1\to s_W)}{P(\bar w;s_{W}\to s_1)}=
\frac{1-P_e(w)}{1-P_e(\bar w)}
\frac{A^{E_2}_{s_2}\ldots A^{E_W}_{s_W}}
{A^{-E_{W-1}}_{s_{W-1}}\ldots
A^{-E_{1}}_{s_{1}}}
\frac{\bar N_{s_w}}{N_{s_1}}
\end{equation}
Since $A^{E_i}_{s_i}/A^{-E_i}_{s_i}=\exp(-\Delta E_i/k_BT)$ we see, using our definition of
$P_e$, that 
\begin{equation}
\frac{P(w;s_1\to s_W)}{P(\bar w;s_{W}\to s_1)}=
\exp(-\Delta E/k_BT).
\label{eq:semidetailE}
\end{equation}
Quite generally, there is more than one way to
introduce a worm to go from $\mu$ to $\nu$.
In fact there are 2 ways of introducing a worm taking the system from
configuration $\mu$ to $\nu$, one where the worm traverses the sites
from $s_1$ to $s_W$ and another where the worm traverses the sites from
$s_W$ to $s_1$. Equivalently, there are 2 ways of introducing the
anti-worm. Hence, $P(\mu\to\nu)$ is a sum of the
two probabilities, $P(\mu\to\nu)=P(w; s_1\to s_W)+P(w; s_W\to s_1)$.
However, employing the result we have just proven we see that:
\begin{eqnarray}
P(\mu\to\nu)= &&\left[P(\bar w; s_W\to s_1)+P(\bar w; s_1\to
s_W)\right]\nonumber\\
&&\times\exp(-\Delta E/k_BT)\nonumber\\
=&&P(\nu\to\mu)\exp(-\Delta E/k_BT).
\end{eqnarray}
Hence, the proposed
algorithm satisfies detailed balance on a bi-partite lattice and
we see that detailed balance is satisfied no matter how many
possible ways one can introduce a worm in order to go from $\mu\to\nu$
as long as the number of possible anti-worms matches. 
Strictly speaking, since the worm can partly erase itself, there {\it is}
in fact an infinite number of worm-moves that will take us from $\mu$ to $\nu$,
however, in each case there is a matching anti-worm so we can replace the above sum
over two probabilities with an infinite sum, yielding the same conclusion.
Ergodicity is
proven by noting that single spin flips are incorporated in the
algorithm.  $\blacksquare$

\section{Proof of Algorithm B\label{app:b}}
The proof is quite similar to the proof of algorithm A and we use the same notation.
The probability for creating
a worm, $w$, starting from site $s_1$ is then given by:
\begin{eqnarray}
& &P(w;s_1\to s_W)=P(W)
P(s_1)\frac{A^{E_2}_{s_2}}{N_{s_1}}
p_{s_2}(s_3|s_1)\ldots\ldots\nonumber\\
& &p_{s_{W-2}}(s_{W-1}|s_{W-3})
p_{s_{W-1}}(s_W|s_{W-2}).
(1-P_e(w))
\end{eqnarray}
Here $P(s_1)$ denotes the probability for choosing site $s_1$ as the
starting site for the worm, $P(W)$ the probability for choosing a worm
length $W$.
$p_{s_i}(s_{i+1}|s_{i-1})$ denotes
the conditional probability, extracted from the minimized matrix $P$, at
the site $s_i$ for continuing to the site $s_{i+1}$ knowing that the
worm is coming from the site $s_{i-1}$.
Starting from the configuration $\nu$ we can now calculate the
probability for introducing an anti-worm, $\bar w$, starting at the site
$s_W$, erasing the worm, $w$. We find:
\begin{eqnarray}
P(\bar w; &s_W\to s_1&)=P(W)
P(s_W)\frac{A^{-E_{W-1}}_{s_{W-1}}}{\bar N_{s_W}}
\bar p_{s_{W-1}}(s_{W-2}|s_W)\nonumber\\
&\ldots&
\bar p_{s_{3}}(s_{2}|s_{4})
\bar p_{s_{2}}(s_1|s_{3})
(1-P_e(\bar w)).
\end{eqnarray}
At a given site, $s_i$, the configurations of the spins during the
construction of the worm and the anti-worm only differ by the position
of the spin at the site itself. Since the set of neighbors \mathversion{bold}$\sigma$\mathversion{normal}
exclude any nearest neighbors to $s_i$ we see that $P_{s_i}=\bar
P_{s_i}$. Hence, by construction, 
\begin{equation}
\frac{p_{s_i}(s_{i+1}|s_{i-1})}{\bar p_{s_i}(s_{i-1}|s_{i+1})}
=
\frac{A_{s_{i+1}}^{E_{i+1}}}{A^{-E_{i-1}}_{s_{i-1}}}.
\end{equation}
Using the definition
of $P_e$ and the fact that $P(s_1)=P(s_W)$, we then see that:
\begin{equation}
\frac{P(w;s_1\to s_W)}{P(\bar w;s_W\to s_1)}=
\frac{A^{E_1}_{s_1}\ldots A^{E_W}_{s_W}}{A^{-E_1}_{s_1}\ldots A^{-E_W}_{s_W}}.
\end{equation}
Since $A^{E_i}_{s_i}/A^{-E_i}_{s_i}=\exp(-\Delta E_i/k_BT)$ we finally see
that 
\begin{equation}
\frac{P(w;s_1\to s_W)}{P(\bar w; s_W\to s_1)}=\exp(-\Delta E/k_BT),
\end{equation}
where $\Delta E$ is the {\it total} energy cost in going from
configuration $\mu$ to configuration $\nu$. 
As was the case for algorithm A, there are 2 ways of introducing a worm taking the system from
configuration $\mu$ to $\nu$ (modulo sites visited an even number of times),
one where the worm traverses the sites
from $s_1$ to $s_W$ and another where the worm traverses the sites from
$s_W$ to $s_1$. Equivalently, there are 2 ways of introducing the
anti-worm. We again see that detailed balance
is satisfied. 
Ergodicity is
proven by noting that single spin flips are incorporated in the algorithm, since for $W=1$
the algorithm corresponds to attempting a Metropolis spin-flip at the selected site.$\blacksquare$

\bibliography{rising}

\end{document}